# The stable phases of the $Cs_2CuCl_{4-x}Br_x$ mixed systems

N. Krüger, S. Belz, F. Schossau, A.A. Haghighirad, P. T. Cong, B. Wolf, S. Gottlieb-Schoenmeyer[1], F. Ritter, W. Assmus

Physikalisches Institut, Goethe-Universität Frankfurt, Max-von-Laue-Strasse 1, 60438 Frankfurt am Main, Germany

[1] Physik-Department E21, Technische Universität München, James-Franck-Str.1, 85748 Garching

**Abstract**

Starting from $Cs_2CuCl_4$ and $Cs_2CuBr_4$ our project focuses on the growth of the $Cs_2CuCl_{4-x}Br_x$ mixed crystals from aqueous solution and the investigation of the occurring structural variations. The well known orthorhombic structure (space group Pnma) of the end members of this system is interrupted within the intermediate composition range $Cs_2CuCl_3Br_1$ - $Cs_2CuCl_2Br_2$, if the growth takes place at room temperature. Within this range a new tetragonal phase is found (space group I4/mmm). However, in case the growth temperature will be increased to 50 °C, the existence of the orthorhombic structure can be extended over the whole $Cs_2CuCl_{4-x}Br_x$ mixed system. A detailed analysis of the composition dependence of the lattice parameters is used to draw conclusions on the incorporation of Cl- and Br-ions at different sites which is important for the magnetic interactions between the Cu-ions.

## Introduction

Within recent years $Cs_2CuCl_4$ attracted a lot of interest, because it was one of the first systems that permitted the observation of field induced Bose-Einstein-condensation of magnons (1). On the contrary, for the iso-structural $Cs_2CuBr_4$ magnon crystallisation is found instead of magnon condensation (2, 3).

The unusual properties of these cesium halogen cuprates are caused by the special arrangement of the constituent $CuX_4$ (X = Cl, Br) tetrahedra and Cs-ions, resulting in a characteristic hierarchy of magnetic interactions:

Along the b-axis the tetrahedra are forming chains providing the predominant antiferromagnetic exchange interaction using the exchange path Cu-Cl-Cl-Cu. The chains are stacked along the c-axis with neighbouring chains being displaced with respect to each other by b/2 giving a two-dimensional frustrated triangular lattice of spin 1/2 states (Figure 1) (4, 5). These layers, which are oriented parallel to the b-c-plane, in turn are stacked along the a-axis with Cs atoms located in between.

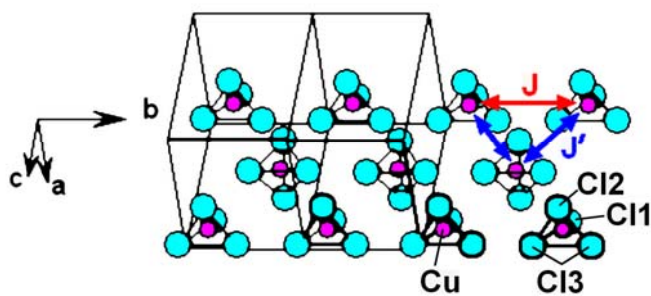

**Figure 1.** Frustrated triangular spin lattice, formed by $CuCl_4$ tetrahedra. The predominant exchange interaction J is realized along b. The ratio J/J´ is about 1/3. The crystallographic different halogen positions are denoted as Cl1, Cl2 and Cl3.

The systematic substitution of Br for Cl according to $Cs_2CuCl_{4-x}Br_x$, $0 \leq x \leq 4$ is regarded as an unique possibility to study the transition from triplet condensation of magnons ($Cs_2CuCl_4$) to triplet crystallisation ($Cs_2CuBr_4$). Ono et al. prepared mixed crystals using melt growth methods to investigate the influence of Cl doping on the magnetic low temperature properties of $Cs_2CuBr_4$ (6). They found that already small amounts of Cl on the Br-sites lead to the suppression of the magnetic ordering. However no information is given about the influence of the dopant on the structural parameters of the investigated compounds.

The aim of this crystal growth project is to provide single crystals over the whole concentration range of the mixed system together with the structural information that is necessary to bring the crossover of the magnetic properties in connection with the variation of the structural parameters.

The composition variation is realised by crystal growth from aqueous solutions, because this method is well established for each of the end members $Cs_2CuCl_4$ and $Cs_2CuBr_4$. This growth method is maintained in order to ensure that the variation of the material properties with respect to the well characterized end members of the mixed crystal series occurs as far as possible continuously.

**Experimental procedure**

The required solute compositions for the growth of $Cs_2CuCl_{4-x}Br_x$ were prepared using the following reaction equations:

$CuCl_2 \cdot 2H_2O + (2-x) \cdot CsCl + x \cdot CsBr \rightarrow Cs_2CuCl_{4-x}Br_x + 2H_2O$

for the concentration range $0 \leq x \leq 2$, and

$CuBr_2 + (4-x) \cdot CsCl + (x-2) \cdot CsBr \rightarrow Cs_2CuCl_{4-x}Br_x$

for the concentration range $2 \leq x \leq 4$.

This implies that with respect to the salts CsX and $CuX_2$ (X = Cl, Br) the stoichiometric mixture 2:1 was used as starting composition for the growth. For the pure end member $Cs_2CuCl_4$, Vasilèva et al. have shown, that solutions with an excess of CsCl are preferable to ensure that this compound is formed as primary phase (7). For the mixed system such comprehensive phase diagram data were not available, but within the whole composition range good quality samples of $Cs_2CuCl_{4-x}Br_x$ could be obtained from stoichiometric mixtures.

The salts used for the experiments are CsCl (≥ 99.999%), CsBr (ultra-clean, VWR), $CuCl_2 \cdot 2H_2O$ (Analar Normapur, Merck) and $CuBr_2$ (≥ 98%).

The mass ratio between the salt mixture and water for preparing the starting solutions was 2:1.

Crystal growth from these aqueous solutions was realised by the evaporation method. Evaporation was preferred over temperature reduction for initiating and maintaining the growth process. As it is shown below the growth temperature is the crucial parameter to select the structural modification for a given mixed crystal composition. Therefore, the whole growth run for a given sample should be carried out at a constant temperature.

During the growth, the solution is kept within a cylindrical polytetrafluoroethylene (PTFE) laboratory beaker (capacity 100 ml, diameter 50 mm) covered with a lid that has a central hole. The hole size is crucial for adjusting the evaporation rate to reach appropriate crystallisation rates. Using lids with large holes or growth from open beakers results in spurious nucleation and eventually in poor crystal quality. The samples for this investigation of the stability ranges of the mixed crystal system have been grown at three different temperature levels. Using lids with hole sizes of one to two centimetres led to three typical durations of the growth process that resulted in crystals with typical edge lengths between 5 and 10 mm, which in most cases was sufficient for sample characterization. The aspect ratio proved to be dependent not only on the growth velocity but also on sample composition and of course on the space group symmetry (Figures 4(a), 4(b)).

1. room temperature: crystallisation within 3 – 4 weeks
2. 50 °C: crystallisation within 2 weeks
3. 8 °C: crystallisation within 6 – 8 weeks

By using lids with a much smaller opening (one to two mm) and growth durations of several months, good quality samples of several cm size can be obtained.

The growth experiments at 8 °C and 50 °C were performed using a temperature controlled testing chamber (Heraeus BK6160).

The thermal stability of these crystals was investigated by means of differential thermal analysis (DTA) and thermogravimetry (TG) using a NETZSCH STA 409 system. For structural characterization, powder x-ray diffractometry (Siemens D500 diffractometer, $Cu_{K\alpha}$-radiation) was used as well as polarisation microscopy. The compositional analysis of the samples was carried out by EDX

## Results

**Growth temperature and phase formation.** Within the whole composition range the halogens are incorporated in the crystal nearly congruently. The differences between the Br/Cl ratios of the solutions and the grown crystals were found to be within the uncertainty range of the EDX analysis. Therefore the data presented here are labelled with the corresponding starting compositions of the growth solutions.

Irrespective of the growth temperature the well known yellow colour of the pure $Cs_2CuCl_4$ is changed to dark red, when Br ions are incorporated at the halogen sites. As a consequence thick samples have a black appearance.

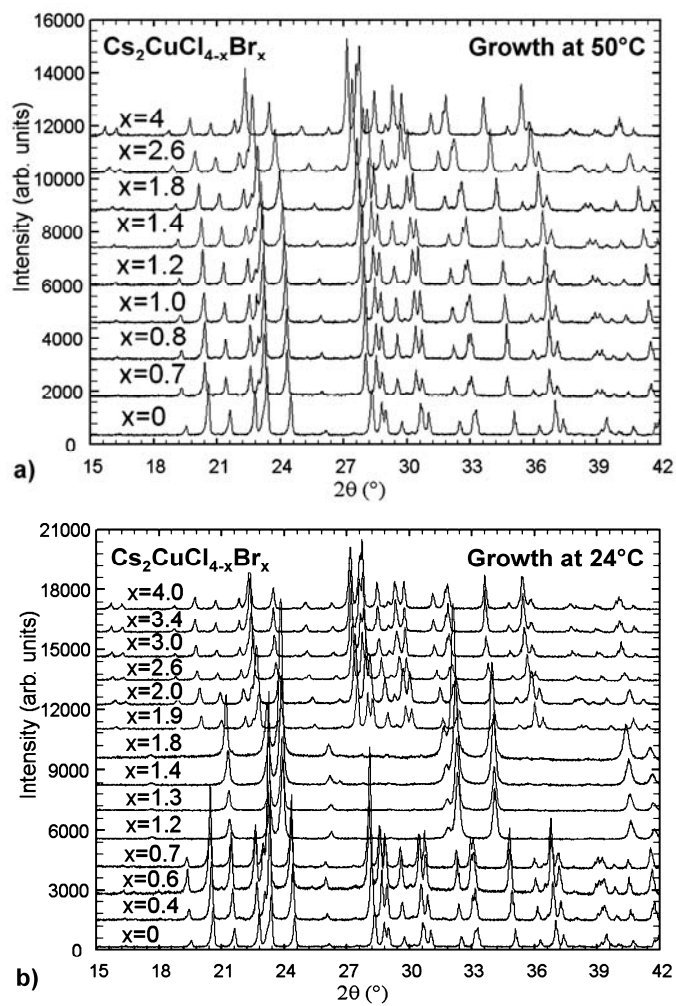

**Figure 2.** Powder XRD on $Cs_2CuCl_{4-x}Br_x$ with different compositions ($Cu_{K\alpha}$-radiation): (a) Growth at 50 °C: The orthorhombic structure (Pnma) is valid for the whole composition range, (b) Growth at 24 °C: The orthorhombic structure (Pnma) is interrupted for $1 \leq x < 1.9$ by a tetragonal structure type (I4/mmm).

For growth temperatures of at least 50 °C, the orthorhombic structure of the end members of the mixed crystal series is preserved over the whole composition range, as can be seen in Figure 2(a), which shows a series of powder diffractograms for $Cs_2CuCl_{4-x}Br_x$ with different Br contents. Vegard's law gives a good approximation for the variation of the cell volume with the Br content (Figure 3(a)).

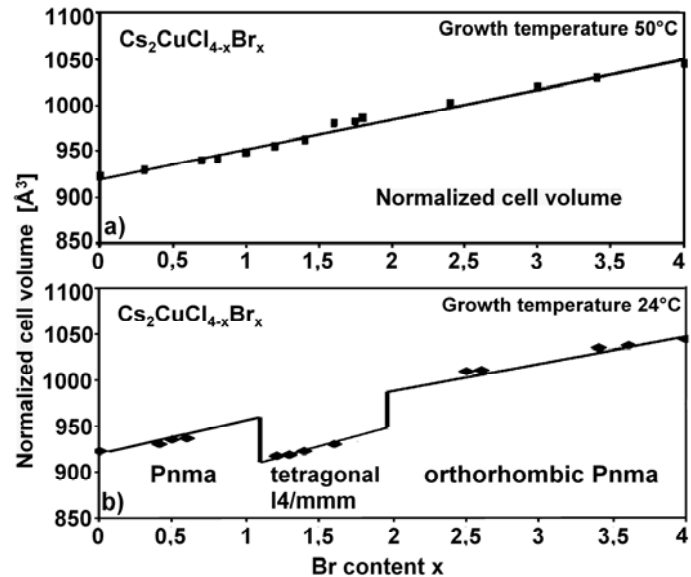

**Figure 3.** Normalized cell volume vs. composition: (a) Growth at 50 °C: Volume expansion with increasing Br-content following Vegard´s law, (b) Growth at 24 °C: Volume contraction within the existence range of the tetragonal modification.

Figure 4(a) shows a representative of these mixed crystals that is typical for a growth duration of one to two weeks.

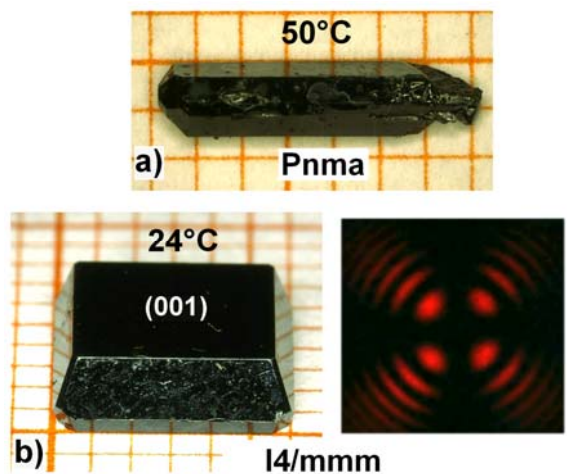

**Figure 4.** Typical habit of $Cs_2CuCl_4$ mixed crystals: (a) Orthorhombic (Pnma) $Cs_2CuCl_{2.4}Br_{1.6}$, grown at 50 °C, (b) Left: Tetragonal $Cs_2CuCl_{2.4}Br_{1.6}$, grown at 24 °C. Right: Conoscopic interference pattern, optical axis parallel to [001].

The structural parameters of the mixed crystal series have been refined (Tables 1 and 2) using the GSAS suite of Rietveld programs (8). As an example the Rietveld plots for two selected crystal compositions are shown in Figures 5(a) and 5(b).

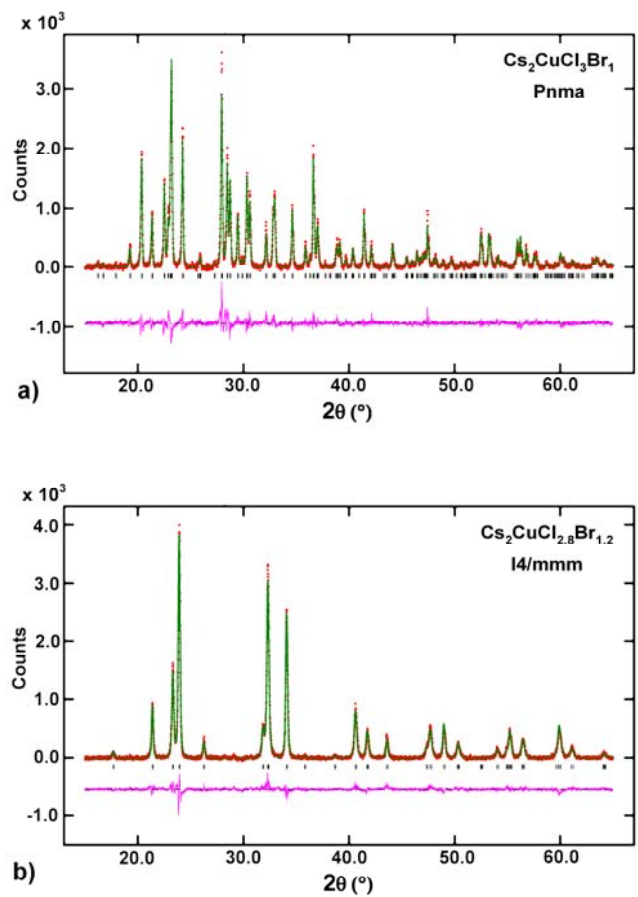

**Figure 5.** Rietveld difference plots for two selected representatives belonging to the orthorhombic ((a), Pnma) and the tetragonal ((b), I4/mmm) structure modification, respectively. The green solid line shows the calculated reflex profile for the refined structure. The difference between the simulation and the measured data is shown by the lower curve. The expected angular positions for the reflections are marked by vertical bars.

The refinement procedure for this large series of diffractograms was carried out using the simplifying assumption that there are no vacancies, i.e. full occupation of all crystallographic sites was assumed. As a consequence corresponding constraints had to be imposed on the fractional occupation parameters of Cl and Br. No constraints have been applied concerning the atomic positions of these halogen constituents, since due to their different ionic radii it was considered reasonable to allow for slightly different coordinates. Tentatively applied constrains on the positions resulted in an increase of the displacement parameters.

For the orthorhombic crystals within the whole composition range the obtained atomic positions come very close to the structural data published by Bailleul et al. for the pure $Cs_2CuCl_4$ (9). As can be seen from Table 1 and Figure 1, the $CuX_4$ (X = Cl, Br) tetrahedra are offering three non equivalent sites to be shared by Cl and Br. The apex of the constituent tetrahedron protruding out of the triangular spin lattice plane is defined by Cl2 and Br2, respectively, occupying the 4c Wyckoff position, while the other fourfold position (4c, Cl1 and Br1), as well as the eightfold one (8d, Cl3 and Br3) refer to the remaining three halogen atoms mediating the exchange within these planes. From the 8d position all these halogen sites are generated, which are aligned along the b-direction (i.e. along the magnetic chains). Concerning the orthorhombic structure the halogen substitution was allowed to take place on all these three Cl sites. This simplification was chosen because this refinement procedure was applied to the whole series of diffractograms.

| $Cs_2CuCl_{4-x}Br_x$, orthorhombic modification, P n m a | | | | | | | | | | | | |
|---|---|---|---|---|---|---|---|---|---|---|---|---|
| | $Cs_2CuCl_4$ | | | $Cs_2CuCl_{3.2}Br_{0.8}$ | | | $Cs_2CuCl_1Br_3$ | | | $Cs_2CuBr_4$ | | |
| a [Å] | 9.753(3) | | | 9.859(2) | | | 10.115(2) | | | 10.170(2) | | |
| b [Å] | 7.609(2) | | | 7.640(1) | | | 7.868(2) | | | 7.956(1) | | |
| c [Å] | 12.394(4) | | | 12.491(2) | | | 12.818(4) | | | 12.915(3) | | |
| Atom | x | y | z | x | y | z | x | y | z | x | y | z |
| Cs 4c | 0.13(3) | 0.25 | 0.10(2) | 0.13(3) | 0.25 | 0.10(3) | 0.13(0) | 0.25 | 0.10(5) | 0.13(1) | 0.25 | 0.10(7) |
| Cs 4c | -0.00(4) | 0.75 | 0.32(7) | -0.00(2) | 0.75 | 0.32(4) | -0.00(1) | 0.75 | 0.32(8) | -0.00(5) | 0.75 | 0.33(0) |
| Cu 4c | 0.23(2) | 0.25 | 0.41(9) | 0.23(3) | 0.25 | 0.41(6) | 0.23(9) | 0.25 | 0.41(5) | 0.23(1) | 0.25 | 0.41(9) |
| Cl1 4c | 0.34(2) | 0.25 | 0.57(1) | 0.34(3) | 0.25 | 0.57(4) | 0.34(6) | 0.25 | 0.57(9) | | | |
| Cl2 4c | 0.01(2) | 0.25 | 0.38(7) | -0.00(7) | 0.25 | 0.36(1) | -0.00(1) | 0.25 | 0.37(2) | | | |
| Cl3 8d | 0.29(8) | -0.00(5) | 0.35(3) | 0.29(7) | -0.00(1) | 0.35(5) | 0.29(3) | -0.01(1) | 0.37(5) | | | |
| Br1 4c | | | | 0.35(1) | 0.25 | 0.57(1) | 0.35(0) | 0.25 | 0.57(4) | 0.34(5) | 0.25 | 0.57(9) |
| Br2 4c | | | | 0.00(2) | 0.25 | 0.38(3) | 0.00(4) | 0.25 | 0.38(1) | 0.00(3) | 0.25 | 0.38(5) |
| Br3 8d | | | | 0.29(5) | -0.00(1) | 0.34(2) | 0.29(8) | -0.01(7) | 0.34(2) | 0.29(6) | -0.01(2) | 0.35(2) |
| $\chi^2_{red}$ | 1.68 | | | 2.87 | | | 2.31 | | | 2.44 | | |
| wRp | 0.056 | | | 0.085 | | | 0.080 | | | 0.079 | | |
| Rp | 0.045 | | | 0.065 | | | 0.063 | | | 0.061 | | |

**Table 1.** Lattice parameters und atomic positions for orthorhombic $Cs_2CuCl_{4-x}Br_x$

If the growth of $Cs_2CuCl_{4-x}Br_x$ takes place at room temperature, within an intermediate concentration range no crystals belonging to the orthorhombic structure type of $Cs_2CuCl_4$ and $Cs_2CuBr_4$ are formed, but instead a new tetragonal phase is obtained. The existence range of the tetragonal modification can be schematically located between $1 \leq x < 2$ with respect to $Cs_2CuCl_{4-x}Br_x$. The exact limits of this range seem to be slightly dependent on the growth temperature. The existence of this tetragonal structure type within these limits is made clear by the crystal habit and the typical conoscopic interference pattern (Figure 4(b)) as well as by the series of powder diffractograms taken for different Br concentrations (Figure 2(b)). By refining the diffraction data the intermediate phase could be assigned to the $K_2NiF_4$ structure type with the space group I4/mmm, that was published earlier by Siefert and Klatyk for $Cs_2CrCl_4$ (10).

The observed tetragonal phase for the intermediate composition range of this mixed system is not related to the tetragonal phase $Cs_2CuCl_4 \cdot 2H_2O$ well known in literature (11, 12) to be obtained from aqueous solution instead of water free orthorhombic $Cs_2CuCl_4$, if the growth temperature is too low (7,13). The diffraction data of our tetragonal mixed crystals cannot be refined using the space group P42/mnm and the cell parameters published for this hydrated phase.

Lattice parameters and atomic positions for different compositions are given in Table 2.

| Cs$_2$CuCl$_{4-x}$Br$_x$ , tetragonal modification, I4/mmm | | | | | | | | | |
|---|---|---|---|---|---|---|---|---|---|
| | Cs$_2$CuCl$_{2.8}$Br$_{1.2}$ | | | Cs$_2$CuCl$_{2.6}$Br$_{1.4}$ | | | Cs$_2$CuCl$_{2.2}$Br$_{1.8}$ | | |
| **a=b** [Å] | 5.2572(6) | | | 5.2651(7) | | | 5.2745(7) | | |
| **c** [Å] | 16.605(4) | | | 16.660(4) | | | 16.735(4) | | |
| **Atom** | x | y | z | x | y | z | x | y | z |
| **Cs** 4e | 0 | 0 | 0.36(2) | 0 | 0 | 0.36(3) | 0 | 0 | 0.36(4) |
| **Cu** 2a | 0 | 0 | 0 | 0 | 0 | 0 | 0 | 0 | 0 |
| **Cl** 4c | 0.5 | 0 | 0 | 0.5 | 0 | 0 | 0.5 | 0 | 0 |
| **Cl** 4e | 0 | 0 | 0.14(5) | 0 | 0 | 0.14(6) | 0 | 0 | 0.14(1) |
| **Br** 4e | 0 | 0 | 0.14(7) | 0 | 0 | 0.14(7) | 0 | 0 | 0.14(7) |
| $\chi^2_{red}$ | 1.72 | | | 2.94 | | | 2.8 | | |
| **wRp** | 0.077 | | | 0.08 | | | 0.06 | | |
| **Rp** | 0.061 | | | 0.063 | | | 0.047 | | |

**Table 2**. Lattice parameters und atomic positions for tetragonal Cs$_2$CuCl$_{4-x}$Br$_x$

This structure consists of CuX$_6$ octahedra (X = Cl, Br), connected to each other by Cs ions. The diffraction data indicate that the Br-dopant is occupying preferably the apex of the CuX$_6$ octahedra resulting in two-dimensional networks of Cu- and Cl-atoms separated from each other by two layers with the composition CsX (X = Cl, Br).

For Cu$^{2+}$ ions in octahedral coordination one has to expect strong Jahn-Teller distortions with not only tetragonal but also orthorhombic symmetry components. Within the framework of this study the centrosymmetric I4/mmm K$_2$NiF$_4$ structure proved to be well suited to describe the diffraction data for the tetragonal or quasi tetragonal modification of the Cs$_2$CuCl$_{4-x}$Br$_x$ mixed crystals. However it has been shown years ago (14, 15, 16) for K$_2$CuF$_4$, that it depends on the degree of ordering of the distorted octahedra, whether I/4mmm or the closely related non centrosymmetric tetragonal (I-4c2) or orthorhombic (Bbcm) space groups give the best approximations of experimentally determined diffraction patterns. In the case of the latter material the actual orthorhombic symmetry was effectively hidden behind multidomain structures (17, 18).

Whether a similar situation is given in the case of our mixed crystal system has to be answered by additional structural characterization.

The tetragonal and the orthorhombic structure types of the mixed crystal system Cs$_2$CuCl$_{4-x}$Br$_x$ cannot be transferred into each other continuously. This can be seen from Figure 3(b) showing the normalized cell volume as a function of the crystal composition. The linear relation according to Vegard`s law is interrupted by the existence range of the tetragonal modification that shows a significant volume contraction with respect to the orthorhombic structure type.

It is observed, that growth of this mixed system at lower temperatures (8 °C in this work) results in an extended composition width for the tetragonal structure type.

**Thermal stability**

**Tetragonal structure type.** The tetragonal phase has limited thermal stability. Heating of $Cs_2CuCl_{2.4}Br_{1.6}$ single crystals up to 200 °C (Figure 6) using a DTA apparatus shows a very clear endothermic signal, indicating a first order transition. On cooling, the corresponding exothermic signal is missing, indicative of an irreversible phase transition.

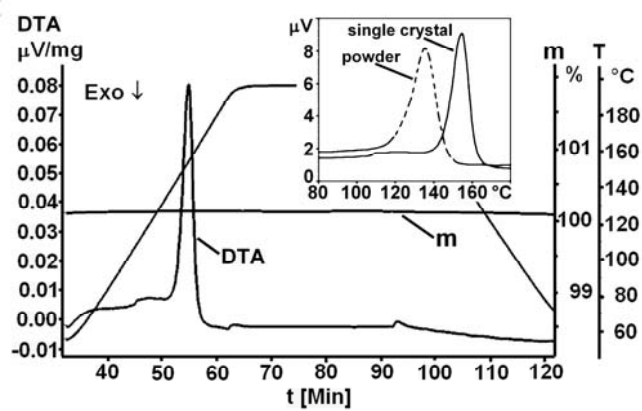

**Figure 6.** Simultaneous differential thermal analysis (DTA) and thermogravimetry (TG) on tetragonal $Cs_2CuCl_{2.4}Br_{1.6}$. Irreversible transformation to the orthorhombic structure type. No mass change is observed.

Subsequent powder diffraction experiments (Figure 7(a)) show, that the DTA peak (Figure 6) can be attributed to the transformation into the orthorhombic structure type, which can be obtained directly for growth temperatures above 50 °C.

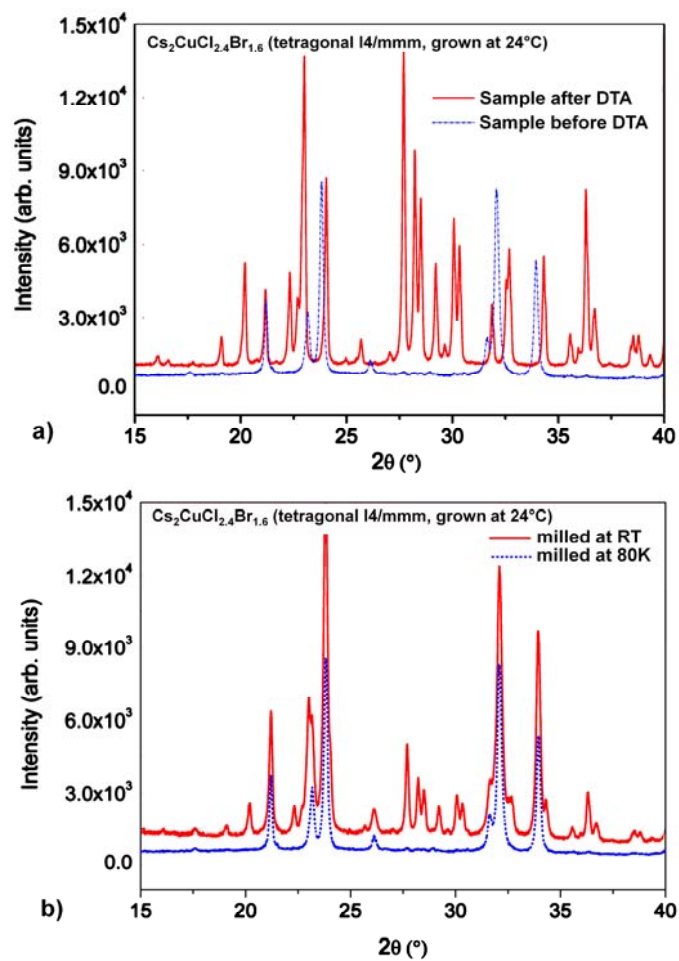

**Figure 7.** Phase transformation of tetragonal $Cs_2CuCl_{2.4}Br_{1.6}$, documented by powder XRD ($Cu_{K\alpha}$-radiation): (a) Phase transition from Pnma to I4/mmm, thermally activated (DTA), (b) Partial phase transition from Pnma to I4/mmm, activated by milling at RT.

Closer inspection shows that the DTA signal of this phase transformation consists of two parts: A step like signal with an onset temperature of around 110 °C is followed by a relatively sharp peak at about 145 °C. Using an optical microscope it can be seen that the whole crystal is broken into very small pieces after DTA treatment and shows a mosaic-like appearance.

To answer the question whether the mechanical destruction of the single crystals is reflected by the DTA peak, the DTA experiment is repeated with powdered tetragonal material. In that case the peak onset temperature is detected near the onset of the step like forepeak mentioned above, but now the DTA signal is not split into two parts. Compared with the single crystal experiment the transition is completed at a lower temperature (see the inset of Figure 6). Thus we can conclude that the shape of the DTA peak could be assigned to the activation energy for the crack formation that is connected to this volume change during the phase transition.

The simultaneously recorded TG signal shows no anomaly at the transition and gives clear confirmation that our tetragonal modification of the $Cs_2CuCl_{4-x}Br_x$ mixed crstals is not related to the hydrated tetragonal phase $Cs_2CuCl_4 \cdot 2H_2O$ mentioned above. For the latter phase a typical weight change of about 7% is observed near 80°C (13).

It has to be mentioned, that the tetragonal-orthorhombic transition is activated not only by heating, but also mechanically by milling at room temperature as it is usually done for preparing powder samples for x-ray diffraction studies. Thinking about the mechanism it seems to be most likely, that a combination of temporarily disturbing the crystal structure and raising the local temperature by the transfer of mechanical energy is responsible for the activation of the phase transition.

The tetragonal phase is not changed, if the powder is milled at low temperatures. In our case the latter was done at about 80 K using a Retsch CryoMill. The formation of the orthorhombic phase as a consequence of the milling process is shown in Figure 7(b).

**Orthorhombic structure type.** Orthorhombic $Cs_2CuCl_{4-x}Br_x$ crystals from the intermediate concentration range ($1 \leq x < 2$), which are only obtained from solutions with temperatures above 50°C, are highly hygroscopic and show significant degradation when they are stored for longer times at room temperature without sufficient protection against humidity. Even if in such crystals no foreign phases are found by diffraction experiments at room temperature, they may show considerable amounts of the tetragonal phase in low temperature experiments and are severely cracked when brought back to room temperature. However no stability problems are encountered with the orthorhombic phase even at very low temperatures, if the crystals are heated shortly to about 150°C prior to use in such experiments.

**Crystal composition and lattice parameters.** As it can be assumed from the already mentioned linear expansion of the cell volume, the lattice constants of the orthorhombic mixed crystal series (grown at T ≥ 50 °C) show in good approximation a linear expansion with increasing Br content (Figure 8(a)). However by closer inspection it is revealed, that depending on the Br concentration range there are slight, but essential anisotropies (Figure 8(b)). For lower dopant concentrations ($0 < x < 1$, range I) the expansion with increasing Br content is found to be the strongest along a-direction that means perpendicular to the triangular spin 1/2 lattice planes, and the weakest along b, the direction of the Cu-X-X-Cu – chains showing the largest exchange interaction. For Br contents higher than 50% ($2 \leq x \leq 4$, range III) we have the opposite situation: The strongest relative expansion is found along the chain direction b, the weakest change occurs perpendicular to the triangular spin lattice planes, i.e. along a.

Within the intermediate composition range ($1 \leq x < 2$, range II) there is no clearly detectable expansion anisotropy. For all three lattice constants (and the volume) the expansion with increasing Br content seems to deviate from linear behaviour in a similarly way. This range appears to be a transition zone between the expansion scenarios observed in the ranges I and III. It is supposed that within this composition range the halogen sites between the spin lattice planes are already completely filled and the additional Br is incorporated into the halogen sites within these planes. The reason for the observable slope change of the lattice parameter expansion within that composition range has to be investigated by high resolution structure investigation.

Taking into account the three non equivalent halogen sites that are permitted by the structure of $Cs_2CuCl_{4-x}Br_x$, one can conclude that for concentration range I ($0 < x < 1$) the Cl2 atoms, that are not part of the spin 1/2 lattice planes are preferably replaced by Br, whereas the substitution of the Cl3 atoms mediating the superexchange within the magnetic chains takes place for higher Br contents (range III, 2 ≤ x ≤ 4).

Although the selective occupation of the halogen sites by Br depending on the concentration range is clearly revealed by the composition dependence of the lattice parameters, it could not be detected unambiguously directly by structure refinement based on an X-ray powder diffractogram due to limited resolution.

However these findings are confirmed by high resolution diffraction data obtained from elastic neutron scattering on single crystals of selected compositions. The complete neutron diffraction study is in preparation (19).

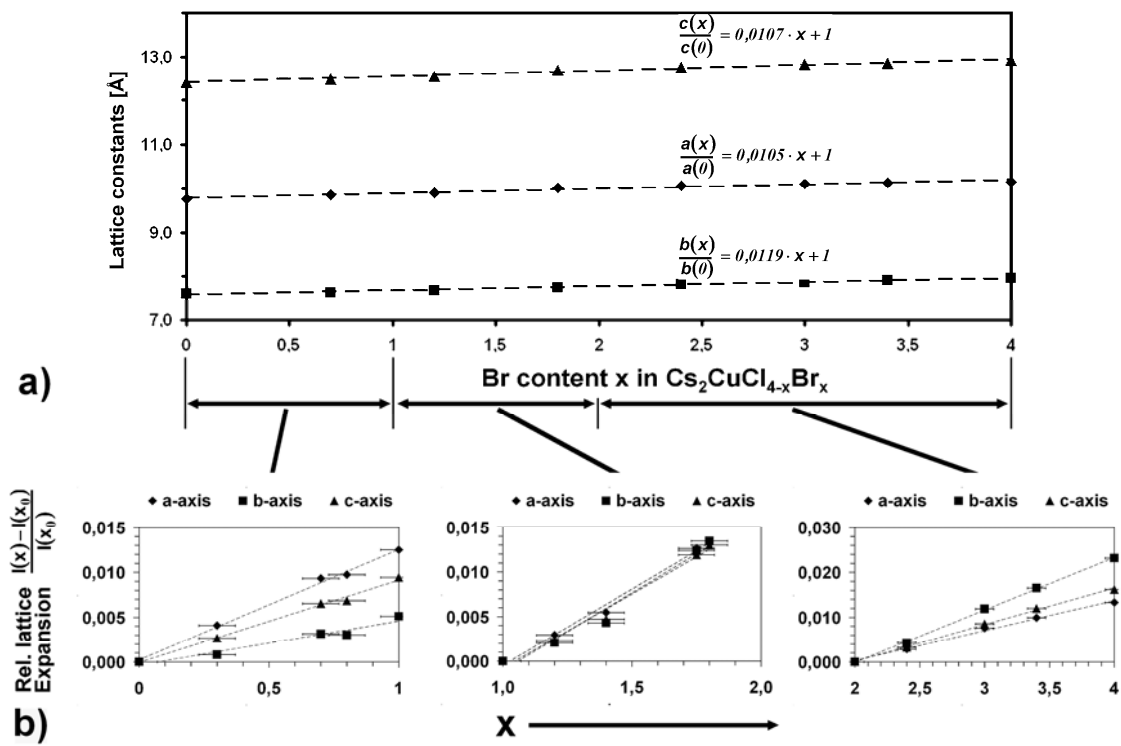

**Figure 8.** Expansion of lattice constants with increasing Br-content: (a) Overview: Nearly linear expansion, indicating mainly that the crystalline structure is unaltered within the whole composition range. (b) The detailed view reveals specific expansion anisotropies for different composition ranges. By relating the change of the lattice constants to their respective starting values the anisotropies of the lattice expansion are made visible. Within the intermediate concentration range there is no anisotropy but deviation from linearity. For each of the three composition ranges the horizontal error bars represent the uncertainties in the EDX results. The vertical statistical error bars are smaller than the data points.

## Conclusions

**$Cs_2CuCl_{4-x}Br_x$ as a model system.** The mixed crystal series $Cs_2CuCl_{4-x}Br_x$ is well suited to serve as a model system for magnon condensation and crystallization phenomena since the orthorhombic structure type exists within the whole composition range. An especially attractive feature of this system is given by the selectivity of the halogen sites that allows the study of the specific influence of a composition variation at one of these sites. As an example the drastic change of the magnetic properties as a result of small amounts of Cl doping on the Br sites, that has been detected by Ono et al. (6), can be qualitatively explained by the fact that for high Br concentrations the halogen substitution affects mainly the chain positions that mediate the strongest exchange interaction within this system.

Of course the overview over the dopant distribution obtained within this work by analysis of the lattice parameters has to be complemented by precise structure analysis that can reveal the dependence between the local and the average crystal structure. Besides the above mentioned single crystal neutron diffraction study (19) a detailed description of the magnetic structure based on the comprehensive investigation of the magnetic properties of this mixed crystal series has reached completion and soon will be submitted for publication (20).

**Phase diagram for the CsX - CuX$_2$ - aqueous solution system and phase stability.** The phase diagram of the CsX - CuX$_2$ - aqueous solution system (X=Cl,Br) leading to mixed crystals Cs$_2$CuCl$_{4-x}$Br$_x$ turns out to be more rich than expected. For the most part of the concentration range more than one structural modification can be obtained, depending on the growth conditions. While for crystals containing both Cl and Br parallel to the orthorhombic structure type (Pnma) a water free tetragonal modification (I4/mmm) is obtained for lower growth temperatures, for the pure Cs$_2$CuCl$_4$ a similar temperature boundary is well known that separates the water free orthorhombic phase (Pnma) from a tetragonal hydrated phase Cs$_2$CuCl$_4$·2H$_2$O.

It seems that for appropriate growth temperatures the stability of the water free tetragonal modification of the mixed crystal system Cs$_2$CuCl$_{4-x}$Br$_x$ can be extended towards the pure end members and further investigations are needed to gain knowledge about the stability ranges of the different structural modifications.

Another question concerns thermal stability of the grown crystals. The tetragonal room temperature modification of the mixed crystal system can be transformed to the orthorhombic structure type by heating, but this phase transition proved to be irreversible. On cooling the orthorhombic mixed crystals have to be well protected against humidity. If there is some moisture on the crystals, they are transformed into the tetragonal phase. Obviously the stabilizing activation threshold which protects the orthorhombic phase from being transferred into the tetragonal modification is significantly lowered by absorbed moisture. The nature of the apparently reconstructive orthorhombic – tetragonal transition should be clarified and the stability limits of the orthorhombic phase under different experimental conditions have to be determined.

**Acknowlegements**

This project was supported by Deutsche Forschungsgemeinschaft SFB/TRR 49.

**Supporting Information**

This information is available free of charge via the Internet at http://pubs.acs.org/.

**Figure Captions**

**Figure 1.** Frustrated triangular spin lattice, formed by $CuCl_4$ tetrahedra. The predominant exchange interaction J is realized along b. The ratio J/J´ is about 1/3. The crystallographic different halogen positions are denoted as Cl1, Cl2 and Cl3.

**Figure 2.** Powder XRD on $Cs_2CuCl_{4-x}Br_x$ with different compositions ($Cu_{K\alpha}$-radiation): (a) Growth at 50 °C: The orthorhombic structure (Pnma) of the end members is valid for the whole composition range, (b) Growth at 24 °C: The orthorhombic structure (Pnma) is interrupted for $1 \leq x < 1.9$ by a tetragonal structure type (I4/mmm).

**Figure 3.** Normalized cell volume vs. composition: (a) Growth at 50 °C: Volume expansion with increasing Br-content following Vegard´s law, (b) Growth at 24 °C: Volume contraction within the existence range of the tetragonal modification.

**Figure 4.** Typical habit of $Cs_2CuCl_4$ mixed crystals: (a) Orthorhombic (Pnma) $Cs_2CuCl_{2.4}Br_{1.6}$, grown at 50 °C, (b) Left: Tetragonal $Cs_2CuCl_{2.4}Br_{1.6}$, grown at 24 °C. Right: Conoscopic interference pattern, optical axis parallel to [001].

**Figure 5.** Rietveld difference plots for two selected representatives belonging to the orthorhombic ((a), Pnma) and the tetragonal ((b), I4/mmm) structure modification, respectively. The green solid line shows the calculated reflex profile for the refined structure. The difference between the simulation and the measured data is shown by the lower curve. The expected angular positions for the reflections are marked by vertical bars.

**Figure 6.** Simultaneous differential thermal analysis (DTA) and thermogravimetry (TG) on tetragonal $Cs_2CuCl_{2.4}Br_{1.6}$. Irreversible transformation to the orthorhombic structure type. No mass change is observed.

**Figure 7.** Phase transformation of tetragonal $Cs_2CuCl_{2.4}Br_{1.6}$, documented by powder XRD ($Cu_{K\alpha}$-radiation): (a) Phase transition from Pnma to I4/mmm, thermally activated (DTA). (b) Partial phase transition from Pnma to I4/mmm, activated by milling at RT.

**Figure 8.** Expansion of lattice constants with increasing Br-content: (a) Overview: Nearly linear expansion, indicating mainly that the crystalline structure is unaltered within the whole composition range  (b) The detailed view reveals specific expansion anisotropies for different composition ranges. By relating the change of the lattice constants to their respective starting values the anisotropies of the lattice expansion are made visible. Within the intermediate concentration range there is no anisotropy but deviation from linearity. For each of the three composition ranges the horizontal error bars

represent the uncertainties in the EDX results. The vertical statistical error bars are smaller than the data points.

**Table 1.** Lattice parameters und atomic positions for orthorhombic $Cs_2CuCl_{4-x}Br_x$

**Table 2.** Lattice parameters und atomic positions for tetragonal $Cs_2CuCl_{4-x}Br_x$

**For Table of Contents Use Only**

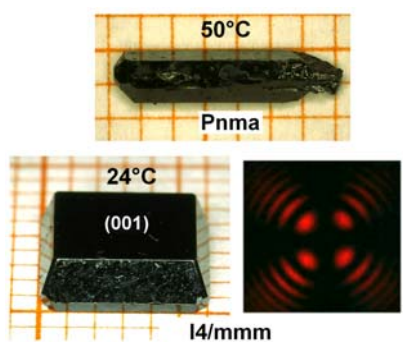

This paper reports on the growth of $Cs_2CuCl_{4-x}Br_x$ mixed crystals from aqueous solutions. Depending on the growth temperature different structural modifications are formed. For T ≥ 50 °C the orthorhombic structure of the frustrated triangular antiferromagnets $Cs_2CuCl_4$ and $Cs_2CuBr_4$ can be obtained for the whole concentration range 0 ≤ x ≤ 4 of the mixed system.